# On the Anisotropy of $E_0 \geq 5.5 \times 10^{19}$ eV Cosmic Rays according to Data of the Pierre Auger Collaboration

**A.V. Glushkov**

a.v.glushkov@ikfia.ysn.ru

*Institute of Cosmophysical Research and Aeronomy, Yakutsk Research Center, Siberian Division, Russian, Academy of Sciences, pr. Lenina 31, Yakutsk, 677980 Russia*

**Abstract**

The Pierre Auger Collaboration discovered, in a solid angle of radius about 18º, a local group of cosmic rays having energies in the region $E_0 \geq 5.5 \times 10^{19}$ eV and coming from the region of the Gen A radio galaxy, whose galactic coordinates are $l_G = 309.5°$ and $b_G = 19.4°$. Near it, there is the Centaur supercluster of galaxies, its galactic coordinates being $l_G = 302.4°$ and $b_G = 21.6°$. It is noteworthy that the Great Attractor, which may have a direct bearing on the observed picture, is also there.

## 1. INTRODUCTION

Data reported in [1] and obtained by the Pierre Auger Collaboration before August 31, 2007, revealed some kind of an anisotropy for 27 cosmic rays above the Greisen–Zatsepin–Kuzmin limit (or GZK cutoff) in the region $E_0 \geq 5.5 \times 10^{19}$ eV. This anisotropy manifests itself as a correlation between a significant fraction of cosmic ray arrival directions and active galactic nuclei (AGN) [2] situated in local volume of space with radius of 75 Mpc. In the new study that was reported by this collaboration in [3], the number of such events was increased to 69. There, a broader class of objects in the surrounding space with a radius of 200 Mpc was considered as possible sources of cosmic rays. These were galaxies from the 2MASS (2 Microns All Sky Survey) catalog [4] and AGN recorded by the Swift Burst Alert Telescope in gamma rays [5]. Particular attention was given to a dense group of cosmic rays constituting 12 events in a solid angle of radius 13º and originating from the region of the Centaurus galaxy supercluster, whose galaxy coordinates are $l_G = 302.4°$ and $b_G = 21.6°$. Below, I present some additional investigations of this group of cosmic rays and considerations concerning its possible nature.

## 2. RESULTS

The arrival direction of 69 cosmic rays from [3] that have energies in the region $E_0 \geq 5.5 \times 10^{19}$ eV and zenith angles in the range $\theta \leq 60°$ are shown in Fig. 1 (closed symbols) in terms of galactic coordinates. Basically, they lie in the Southern Hemisphere of the Earth. For the picture to be clearer and for the clarity of the arguments given below, the arrival directions are also given here for 32 cosmic rays measured at the Yakutsk array for recording extensive air showers over the entire period of its operation till April 2011 (closed crosses) and characterized by energies in the region $E_0 \geq 4.4 \times 10^{19}$ eV and $\theta \leq 60°$. The threshold energy for these events was chosen in such a way that the average densities of cosmic rays in the Southern and Northern Hemispheres were approximately equal.

The density distributions of these events with respect to each other were investigated in the present study. In Fig. 1, the gray tones represent the deviations of the observed number of events, $N_1$, from the expected number, $\langle N \rangle = N_2 (\Omega_1/\Omega_2)$, in units of the standard deviation, $\sigma = \sqrt{\langle N \rangle}$, that is,

$$n_\sigma = (N_1 - <N>)/\sigma, \qquad (1)$$

where $N_1$ and $N_2$ are the numbers of showers in the solid angles $\Omega_1 = 2\pi(1 - \cos\theta_1)$ and $\Omega_2 = 2\pi(1 - \cos\theta_2)$, respectively, with $\theta_1 = 20°$ and $\theta_2 = 90°$. The values of the expression in (1) were determined by moving the center of an area $1°\times 1°$ in size over the entire sphere. The limits within which $n_\sigma$ changes are shown in the lower part of the figure as a toned scale. The lightest tones correspond to a cosmic ray flux excess of $n_\sigma \geq 3\sigma$ above the average level.

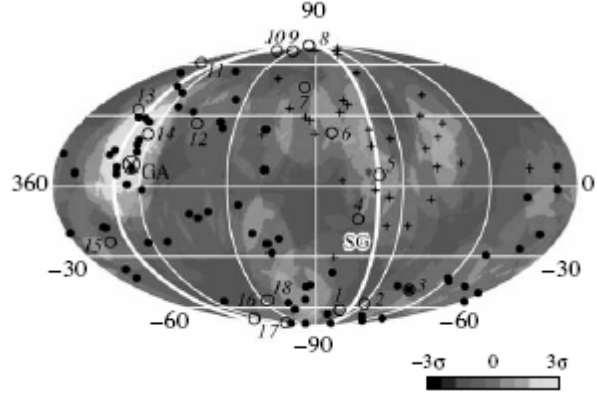

**Fig. 1.** Arrival directions of 69 cosmic rays from [3] with $E_0 \geq 5.5\times 10^{19}$ eV (dark symbols) and 32 cosmic rays with $E_0 \geq 4.4\times 10^{19}$ eV from the Yakutsk array for recording extensive air showers (crosses), the zenith angles of the latter lying in the range of $\theta \leq 60°$, in the developed celestial sphere in terms of galactic coordinates. The open circles represent the centers of the galaxy superclusters shown in Fig. 2, while the dashed circle is the center of Great Attractor (GA). The SG curve stands for the Supergalaxy plane.

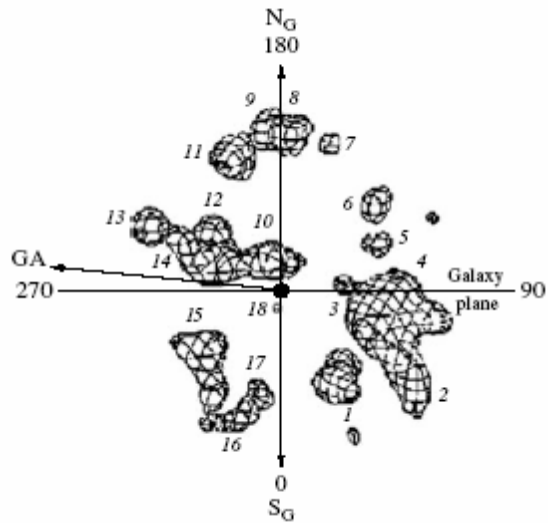

**Fig. 2.** Density distribution of galaxies in a local volume of radius about 100 Mpc in the Supergalaxy plane, the North Pole pointing to the observer [7]. The Supergalaxy longitude is reckoned from the "0" axis in the counterclockwise direction. Here, NG and SG are, respectively, the North and the South Pole of the Galaxy, whose plane is orthogonal of the plane of the figure and coincides with the 90–270 line. The GA arrow is directed toward the Great Attractor. Our position is indicated by the closed circle. The figures stand for the numbers of superclusters whose names are given in the table.

Largest galaxy superclusters in a local space volume of radius about 100 Mpc

| No. | Name of object | Galactic coordinates | |
|---|---|---|---|
| | | $l_G$ | $b_G$ |
| 1 | A 194 | 144.5 | −62.4 |
| 2 | Pegasus 1 | 85.1 | −47.6 |
| 3 | Pisces | 119.0 | −56.8 |
| 4 | Perseus | 150.4 | −13.4 |
| 5 | Camelopardalis | 136.4 | 4.7 |
| 6 | A 569 | 168.3 | 22.0 |
| 7 | A 779 | 191.1 | 44.4 |
| 8 | Coma | 244.3 | 85.5 |
| 9 | A 1367 | 236.0 | 73.4 |
| 10 | Virgo | 279.7 | 74.4 |
| 11 | 7S 224 | 343.6 | 61.8 |
| 12 | Hydra | 269.6 | 26.5 |
| 13 | A 3581 | 323.2 | 32.9 |
| 14 | Centaurus | 302.4 | 21.6 |
| 15 | Pavo | 332.2 | −23.6 |
| 16 | A 2870 | 300.8 | −70.2 |
| 17 | A 2911 | 271.3 | −76.9 |
| 18 | Fornax 1 | 236.3 | −54.6 |
| 19 | Great Attractor | 307.0 | 9.0 |

One can see that the distribution of cosmic ray arrival directions is relatively uniform. This is not so only for 15 events in the region of a light spot (the respective galactic coordinates are $l_G \approx 307°$ and $b_G \approx 9°$) in a solid angle of radius about 18°. On average, there must be about 3.4 such events for the entire sample. The excess is $(15 − 3.4)/\sqrt{3.4} \approx 6.2\sigma$, which is highly improbable to be due to a random coincidence.

In [3], the Pierre Auger Collaboration tried to find sources of cosmic rays having the above energies in the near surrounding extragalactic space. In the opinion of those authors, these may be AGN (Seyfert galaxies) in a local volume of radius 75 Mpc. According to the estimates presented in [3], the probability for the random coincidence of the arrival directions in 69 events with these objects from the catalog in [2] within solid angles of radius 3.1° is $p \approx 0.003$. The fraction of events correlated with AGN is $38^{+7}_{-6}$ %, which is to be compared with the fraction of 21% expected for an isotropic flux of cosmic rays; the excess is at a level of three standard deviations ($3\sigma$). This is smaller than the value of $69^{+11}_{-13}$ % found previously in [1]. We note that the HiRes Collaboration [6] found no correlations in their data with AGN from the catalog presented in [2]. In view of this, the authors of [3] themselves recognize that the question of whether AGN may be sources of cosmic rays whose energies are above the GZK cutoff has remained open to date—to answer it definitively, we need a further substantial increase in the statistics of respective events.

### 3. DISCUSSION

Yet, we will try to extract from the data in Fig. 1 some useful information, which, in the opinion of the present author, may be of interest for further investigations. Figure 2 shows the density distribution of galaxies in a space volume of radius 100 Mpc in the plane of the local galaxy supercluster (Supergalaxy), whose North Pole points to the observer [7]. The Supergalactic longitude is reckoned from the "0" axis in the counterclockwise direction. The NG and SG axes indicate the directions to, respectively, the North and the South Pole of the Milky Way Galaxy, whose plane is orthogonal to the plane of the figure and coincides with the 90–270 line. Our position is indicated by the closed circle. The GA arrow is directed toward the center of the Great Attractor, which will be discussed below.

One can see that galaxies form giant superclusters between which there are voids of enormous size. The names of superclusters and their coordinates are given in the table. The centers of these superclusters are represented by open circles in Fig. 1, the figures there corresponding to the numbers in the table. The overwhelming majority of superclusters form some kind of a flat pancake lying in the supergalaxy plane (light SG curve) – in the supergalaxy latitude region of $|b_{SG}| \leq 30º$. The arrival directions of the majority of cosmic rays in Fig. 1 do not correlate with the positions of superclusters. This is yet another piece of evidence that Seyfert galaxies may hardly be sources of cosmic rays having extremely high energies, since the Seyfert galaxies themselves lie in galaxy superclusters.

It is the opinion of the present author that the aforementioned dense group of 15 events in Fig. 1 deserves particular attention. The authors of [3] indicate that its center coincides with the position of the Gen A radio galaxy in the Centaurus supercluster of galaxies (circle *14*), which may be sources of cosmic rays having extremely high energies. However, we indicate that the Great Attractor lies precisely at this place of the sky. Its center, whose galactic coordinates are $l_{GA} \approx 307º$ and $b_{GA} \approx 9º$, is represented by the dashed circle in Fig. 1. According to the estimates presented in [8], the gravitating mass of the Great Attractor is $M_{GA} \sim 5\times10^{16}\,M_S \approx 10^{47}$ kg (where $M_S$ is the Sun's mass), its center being at a distance of $R_{GA} \approx 58$ Mpc from us. This object of extremely large mass exceeds all known galaxy superclusters by a factor of 30 to 100. We can assume that, near the center of the Great Attractor, there is a dense group of cosmic ray sources, which lead to the observed picture. Possibly, these are the same sources as those that generate the remaining cosmic rays in Fig. 1, but, because of their high concentration in the Great Attractor, they manifest themselves in the form of the observed cosmic ray spot smeared by extragalactic and galactic magnetic fields. No such relation in direction to the remaining galaxy superclusters is observed because of the weakness of these objects as cosmic ray sources.

However, we cannot rule out the possibility that we are dealing here with a manifestation of some other phenomenon. In this connection, we will consider an intriguing fact that will be of use in our subsequent arguments. The results of an investigation of the structure of matter in a volume of radius about 380 Mpc are presented in [9]. The spatial distribution of 535 rich galaxy clusters characterized by red shifts of $z \leq 0.1$ and taken from the catalog of Abell and his coauthors [10] is presented in that article. Figure 3 shows the deviations

$$n_z = (<z> - <z>_0)/\Delta z \qquad (2)$$

of the average red shifts $<z>$ in a circle of radius 45° with respect to the average red shift of $<z>_0 = 0.056 \pm 0.001$ for the entire data sample, the center of this circle being consecutively moved over all areas of the sphere 1°×1° in size. The limits of variations in $n_z$ are shown at the bottom of the figure in the form of a toned scale with a step of $\Delta z = 0.004$. The dark SG curve represents the Supergalaxy plane, while curve *1* stands for the Earth's equator; S and N are, respectively, the South and the North Pole of the Earth.

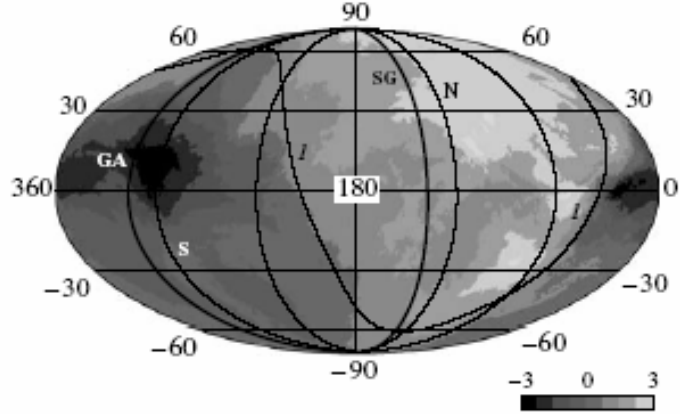

**Fig. 3.** Distribution in the developed celestial sphere in terms of galactic coordinates for the deviation in (2) of the red shifts of 535 rich galaxy superclusters from the catalog in [10] with $z \le 0.1$ from $<z>_0 = 0.056$ for the entire data sample. The toned scale shows the limits of variations in $n_z$ with a step of $\Delta z = 0.004$. Curve *1* stands for the Earth's equator; N and S are, respectively, the North and the South Pole of the Earth; GA is the Great Attractor; and the SG curve is the Supergalaxy plane.

In Fig. 3, an anisotropy of red shifts in the form a black "hole" coinciding with the white spot of the cosmic ray flux in Fig. 1 is clearly seen near the intersection of the Galaxy and Supergalaxy planes. This black "hole" corresponds to minimum red shifts and coincides with the position of the Great Attractor. There are reasons to assume that these two astrophysical phenomena are not random. They are related to each other and to the general distribution of matter in a Metagalaxy – a gravitationally coupled totality of matter in the Universe where we live and beyond which our sight cannot penetrate in principle. A similar "hole" was discovered by the present author [9] in the spatial distribution of Seyfert galaxies and quasars characterized by red shifts of $z \le 6$. The positions of these objects display a global anisotropy caused by a shift of observer's position by approximately 50 Mpc from their symmetry center (presumably Metagalaxy center) toward the vector whose galaxy coordinates are $l_G \approx 123°$ and $b_G \approx 7°$. In the direction opposite to this, there is a vast region where the red shift decreases gradually to a minimum value (in the vicinity of the point whose coordinates are $l_G \approx 303°$ and $b_G \approx -7°$). It is the opinion of the present author that this topological inhomogeneity is one of many indications of the actual existence of a Metagalaxy. The idea of a Metagalaxy as such is not new. From time to time, it is developed by some researchers (see, for example, [9, 11–18]). In [14, 15], it was proposed to consider the Universe as that which consists of a great many closed miniuniverses isolated from one another. Possibly, it is our Metagalaxy that is one of such miniuniverses.

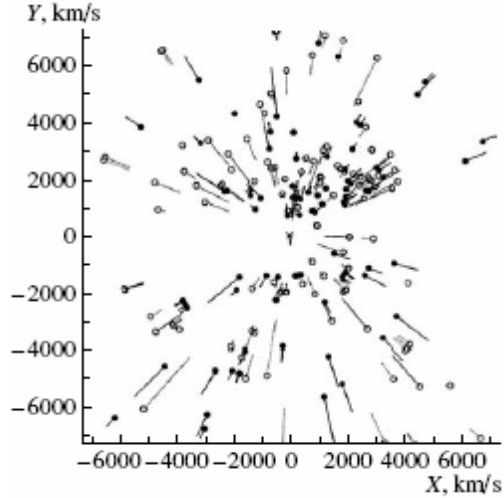

**Fig. 4.** Radial vectors of velocities of individual elliptic galaxies (open circles) and their groups containing not less than three members (closed circles) in a cone of angle 45° with respect to the direction specified by the galactic coordinates $l_G = 307°$ and $b_G = 9°$ (cross at the center of the figure) [8]. The light horizontal band at $Y = 0$ is caused by light absorption in the Galaxy disk.

Anyway, this idea permits, in my opinion, to interpret in some way the results described above. Roughly, the light spot of the excess cosmic ray flux in Fig. 1 corresponds to the center of the Metagalaxy. Under these conditions, the Metagalaxy gravitational potential, which has a spherically symmetric shape, plays an important role in the formation of the direction of motion of cosmic rays (along with their sources). This is manifested, for example, in the collective motion of galaxies belonging to a local volume of space. Figure 4 shows radial vectors of velocities f individual elliptic galaxies (light circles) and their groups containing not less than three members (dark circles) in a cone of angle 45° with respect to the direction specified by the galactic coordinates $l_G = 307°$ and $b_G = 9°$ (cross at the center of the figure). The light horizontal band at $Y = 0$ is caused by light absorption in the Galaxy disk. The motion of the galaxies occurs simultaneously in two directions – to the center of the Great Attractor and away from it. It seems that something like this may also occur to cosmic rays of extremely high energies. According to the analysis of the present author in [11], quasars generating neutral particles may be sources of cosmic rays that have such energies. Some of them pass in the vicinity of the Metagalaxy center and possibly suffer gravitational focusing, which leads to the picture observed in Fig. 1.

## 4. CONCLUSIONS

We have studied the distribution of arrival directions for 69 cosmic rays from [3] that have energies in the region of $E_0 \geq 5.5 \times 10^{19}$ eV and zenith angles in the range of $\theta \leq 60°$. The cosmic rays were recorded by the Pierre Auger Collaboration in the period before December 31, 2009. In Fig. 1, one can clearly see a statistically significant (at a level of $6.2\sigma$) cluster of 15 events in a solid angle of radius about 18°. In addition to the statement of the authors of [3] themselves that the midpoint of this cluster coincides with the position of the Gen A radio galaxy and the position of the Centaurus galaxy supercluster, we have also noticed in the above analysis that the Great Attractor, which is an object of an extremely large mass, also lies here, the respective galactic coordinates being $l_{GA} \approx 307°$ and $b_{GA} \approx 9°$. It exceeds all known galaxy superclusters by a factor of 30 to 100 and may contain an anomalously dense group of cosmic ray sources, which lead to the observed picture. Possibly, the situation around the sources is even more intricate, and it is necessary to consider a much larger volume of surrounding space. This is suggested by a pronounced anisotropy of red shifts of rich galaxy clusters in Fig. 3. According to data from the Yakutsk array for recording extensive air showers, quasars may be sources of cosmic rays having energies in the region of $E_0 \geq 10^{19}$ eV [11]. In this case, those cosmic rays that pass in the vicinity of the Great Attractor may suffer gravitational focusing because of the effect of the Metagalaxy gravitational field and generate the picture observed in Fig. 1. However, there are still many open questions here calling for further investigations.